\documentclass[twocolumn,superscriptaddress,nofootinbib,preprintnumbers,amsmath,amssymb]{revtex4-1}
\usepackage[T1]{fontenc}

\usepackage[titles]{tocloft}
\cftsetindents{section}{0em}{2.5em}
\cftsetindents{subsection}{2.5em}{2.5em}

\usepackage{multirow}
\usepackage{tabularx}
\usepackage{amsfonts}
\usepackage{mathrsfs}
\usepackage{leftidx}
\usepackage{amssymb}
\usepackage{placeins}
\usepackage{relsize}
\usepackage{slashed}

\usepackage[dvipsnames]{xcolor}
\definecolor{red}{rgb}{0.9, 0,0}
\definecolor{cerulean}{rgb}{0., 0.42,0.9}
\definecolor{navy}{rgb}{0.05, 0.05,0.8}

\usepackage[colorlinks]{hyperref}
\hypersetup{
     colorlinks   = true,
     citecolor    = red,
	linkcolor = navy
}

\usepackage{epsfig}
\usepackage{graphicx}               
\usepackage{url}
\usepackage{float}
\usepackage{soul}

\newcommand{\be}{\begin{equation}}
\newcommand{\ee}{\end{equation}}
\newcommand{\bea}{\begin{eqnarray}}
\newcommand{\eea}{\end{eqnarray}}
\newcommand{\beq}{\begin{eqnarray}}
\newcommand{\eeq}{\end{eqnarray}}
\def\bit{\begin{itemize}}
\def\eit{\end{itemize}}
\def\ben{\begin{enumerate}}
\def\een{\end{enumerate}}

\newcommand\DN[1][\relax]{%
\ifx\relax#1\relax\else{}^{#1}\fi \!X}

\DeclareMathAlphabet\mathbfcal{OMS}{cmsy}{b}{n} 

\graphicspath{{./figures/}}

\definecolor{cerulean}{rgb}{0., 0.52,0.65}

\begin{document}


\title{Spin-Dependent Light Dark Matter Constraints from Mediators}

\author{Harikrishnan Ramani}
\affiliation{Berkeley Center for Theoretical Physics, University of California, Berkeley, CA 94720}
\affiliation{Theoretical Physics Group, Lawrence Berkeley National Laboratory, Berkeley, CA 94720}
\author{Graham Woolley}
\affiliation{Department of Physics, University of California, Berkeley, CA 94720}

\begin{abstract} 
A bevy of light dark matter direct detection experiments have been proposed, targeting spin-independent dark matter scattering.  In order to be exhaustive, non-standard signatures that have been investigated in the WIMP window including spin-dependent dark matter scattering also need to be looked into in the light dark matter parameter space. In this work, we promote this endeavor by deriving indirect limits on sub-GeV spin-dependent dark matter through terrestrial and astrophysical limits on the forces that mediate this scattering.

\end{abstract}

\maketitle


\section{Introduction}
Repeated null results in the highly motivated WIMP window have inspired vistas into lighter dark matter below the GeV range, e.g \cite{Battaglieri:2017aum} and references therein. New direct detection (DD) strategies \cite{Knapen:2016cue, Knapen:2017ekk,Griffin:2018bjn,Kurinsky:2019pgb,Abramoff:2019dfb,Hochberg:2019cyy} with lower energy thresholds to target this mass range, as well as experiments to probe the mediator itself \cite{Bondi:2017gul,Akesson:2018vlm,Benato:2018ijc} have been proposed recently. 

Spin-dependent (SD) dark matter models are well motivated in literature \cite{Jungman:1995df,Agrawal:2010fh,Fan:2010gt} and couple only to the nuclear spin. Since most common spin-independent (SI) DM detector targets have an even number of (and hence paired) protons and neutrons, they are not as sensitive to spin-dependent DM. SD interactions have been investigated around the GeV range and above in \cite{Behnke:2016lsk,Amole:2017dex,Agnese:2017jvy,Collar:2018ydf,Bringmann:2018cvk,Fu:2016ega,Akerib:2017kat,Agnese:2017jvy,Aprile:2019dbj} with odd-p or odd-n isotopes. However, all DD proposals for light DM hitherto have explored only the SI case. Intriguingly, since light dark matter detection necessarily involves lighter nuclear targets, the loss in coherence will not be as limiting as in WIMP DM detection. Also, self-interaction constraints, which limit the DM - mediator coupling could also be relaxed if the DM self-interaction involves suppression factors.

Thus, the viability of models of DM that dominantly interact with the SM through spin-dependent couplings below the GeV scale becomes an interesting question. Purely SD interactions are hard to model-build even around the GeV scale\cite{Fan:2010gt, Freytsis:2010ne}. Furthermore, as pointed out in \cite{Green:2017ybv,Knapen:2017ekk}, stringent limits on a light mediator already set severe limits on MeV scale DM albeit it was explored only in the SI context. In this letter, we use a similar strategy to set limits on SD DD cross-section and explore the viability of parameter space to determine if future SD DD experiments are well-motivated.

\section{Models}

\begin{table*}[ht!]
\centering
\begin{tabular}{|c|c|c|c|c|c|c|c|}
\hline
{\bf DM coupling} &{\bf SM coupling} & {\bf mediator }& {\bf Operator} &  \multicolumn{2}{c|}{\bf Direct Detection}& \multicolumn{2}{c|}{\bf Self-Interaction} \\
\cline{5-8}

&&&&  {\bf Massless}&  {\bf Massive} &  {\bf Massless} &  {\bf Massive} \\
\hline
\hline
$\phi^2 a$  & \multirow{4}{*}{$\bar{N}  \gamma_5 N a$} &  \multirow{4}{*}{pseudoscalar} & $O^s_2$  & \multirow{3}{*}{$\frac{1}{q^2} $}&  \multirow{3}{*}{$q^2$}&  \multirow{3}{*}{$\frac{1}{q^4}$} &  \multirow{3}{*}{$1$}\\ 
$\bar{\chi} \chi  a$  &  &  & $\mathbf{O^f_3}$  & &  & &\\ 
$B_{\mu} B^{\mu}  a$  &  &  & $O^v_2$  & & & &\\ 
\cline{1-1}
\cline{4-8}
$\bar{\chi} \gamma_5 \chi  a$  &  &  & $\mathbf{O^f_4}$  & 1 &  $q^4$ & 1 &$q^4$\\ 
\hline
\hline
$\phi^\dagger \partial_\mu \phi A^\mu$  & \multirow{6}{*}{$\bar{N}  \gamma_\mu \gamma_5 N A^{\mu}$} &  \multirow{6}{*}{axial-vector} & $O^s_4$  & \multirow{3}{*}{$\frac{v^2}{q^4}$}&  \multirow{3}{*}{$v^2$}&  \multirow{3}{*}{$\frac{1}{q^4}$} &  \multirow{3}{*}{$1$}\\ 

$\bar{\chi} \gamma_\mu  \chi  A^\mu$  &  &  & $O^f_7$  & & & & \\ 
$B^\dagger_{\nu} \partial_\mu B^{\nu}  A^\mu$  &  &  & $O^v_4$  & & & &\\ 
\cline{1-1}
\cline{4-8}
$B^\dagger_{\nu} \partial_\nu B^{\mu}  A^\mu$  &  & & $\mathbf{O^v_6}$  &$\frac{1}{q^2}$ &$q^2$&$1$ &$q^4$\\ 
\cline{1-1}
\cline{4-8}
$\bar{\chi} \gamma_\mu \gamma_5 \chi  A^\mu$ &  &  & $\mathbf{O^f_8}$  & \multirow{2}{*}{$\frac{1}{q^4}$}&  \multirow{2}{*}{$1$}& \multirow{2}{*}{$\frac{1}{q^4}$}&  \multirow{2}{*}{$1$}\\ 
$\epsilon_{\sigma \nu \rho \mu} B^{\sigma} \partial^\nu B^{\rho}  A^\mu$  &  &  & $O^v_8$  & & & &\\ 
\hline
\hline
$\bar{\chi} \sigma_{\mu\nu}  \chi  F^{\mu \nu}$ & \multirow{4}{*}{$\bar{N} \sigma_{\mu\nu}  N  F^{\mu \nu}$}   & \multirow{4}{*}{vector}  & $O^f_9$  & 1& $q^4$ &1 & $q^4$\\
\cline{1-1}

\cline{4-8}

$\phi^\dagger \partial_\mu \phi A^\mu$   & & &$O^s_{5}$ &  \multirow{3}{*}{$\frac{1}{q^2}$} & \multirow{3}{*}{$q^2$} & \multirow{3}{*}{$\frac{1}{q^4}$}&\multirow{3}{*}{1} \\
$\bar{\chi} \gamma_\mu  \chi  A^\mu$  &  &  & $O^f_{11}$  & & & & \\ 
$B^\dagger_{\nu} \partial_\mu B^{\nu}  A^\mu$  &  &  & $O^v_{9}$  & & & &\\ 

\hline
\end{tabular}
\caption{Operators that produce solely spin-dependent interactions are enumerated. They are broadly organized in terms of the mediator - pseudoscalar, axial-vector or vector through dipole coupling. Also tabulated are the DD cross-section as well as the dark matter self-interaction induced in both the massless and massive limits.}
\label{tab1}
\end{table*}

Purely spin-dependent interactions between DM and SM target are mediated by a pseudoscalar , an axial-vector or a dipole interaction mediated by a dark vector. This mediator can further couple in a plethora of ways with DM.  An exhaustive list of operators was studied in detail in \cite{Fan:2010gt, Freytsis:2010ne} and the operators that produce solely spin-dependent interactions are enumerated in Table.~\ref{tab1}. Added to this are 3 operators that represent DM charge - SM dipole scattering which are momentum dependent. 

The operators are broadly organized in terms of the relevant mediator - pseudoscalar, axial-vector or a vector coupled to nucleons through dipole interactions. Also tabulated are the DD cross-section as well as the dark matter self-interaction induced in both the massless and massive limits. The "operator" term which is typically used only for the massive mediator limits, is used both in the massless and massive limits for brevity.

In the same vein as in \cite{Knapen:2017ekk} the strategy is to minimize self-interaction so as to evade bullet-cluster limits\cite{Kahlhoefer:2013dca} and maximize the DD cross-section so as to set the most conservative limits arising from mediators.

Among the 4 unique options relevant to pseudoscalar mediators, $\mathcal{O}^s_2, \mathcal{O}^f_3, \mathcal{O}^v_2$ have the same cross-section behavior and differ only in the spin of DM. Thus we shall consider only $\mathcal{O}^f_3$ and the rest differ only by $O(1)$ numbers. $\mathcal{O}^f_4$ will be considered in its own right and hence these two operators will be considered. 

Among the axial-vector mediated operators, $\mathcal{O}^f_8, \mathcal{O}^v_8$ have identical cross-section behavior and hence only $\mathcal{O}^f_8$ will be considered. $\mathcal{O}^s_4,~\mathcal{O}^f_7,~\mathcal{O}^v_4$ have the same self-interaction cross-section as $\mathcal{O}^f_8$ but have velocity suppressed DD cross-section and hence will always be more strongly constraining.  $\mathcal{O}^v_6$ is relevant only for vector dark matter and will be considered in its own right. In the SM, Z exchange can generate some of these operators, however for light DM, Z decay poses severe constraints on the coupling and hence the cross-sections. Hence we we will consider only a new axial-vector mediator to generate these interactions.

Finally, the vector dipole coupling to the SM produces $\mathcal{O}^f_9$ and three other operators. While the SM photon is the ideal mediator, it generically produces dipole-dipole as well as dipole-charge and charge-charge interactions between the "milli-charge" DM and the SM, the latter of which is spin-independent\cite{Banks:2010eh}. The relative ratio of SD and SI interactions in the $m_{\chi} \ll m_N$ limit is,  
\begin{equation}
\frac{\sigma_{\rm SD}}{\sigma_{\rm SI}}\sim \frac{ m_\chi ^2}{Z_{\rm eff}^2 m_N^2}
\end{equation}
where $Z_{\rm eff}$ is the effective momentum-exchange-dependent charge DM sees and is given by, 
\begin{equation}
Z_{\rm eff}= \frac{a_0^2 q^2}{1+a_0^2 q^2} Z
\end{equation}
This is Thomas-Fermi screening, with $a_0$ the Bohr radius. Apart from the lack of coherent-enhancement there is also a momentum suppression. However for much smaller DM masses, i.e. $m_{\chi} \ll \frac{1}{a_0 v} \sim 4$ MeV, there is large amount of screening and $Z_{\rm eff} \rightarrow a_0^2 q^2 Z$ and at some mass, DM will only scatter with the SM dipole.
However, for SD to overcome SI one has to go all the way down to $m_{\chi} \sim 13 \text{keV}$ or below. There are strong constraints on milli-charge particles at these masses from stellar constraints. The same statements are true for the vanilla dark photon with kinetic mixing. 

In order to see SD signal first, we need to introduce a new vector that couples only through dipole interactions. The limits on such a force, that couples purely through the dipole with nuclei has not been studied in literature, but in general they should be comparable with the pseudoscalar. Furthermore, the DD cross-section as well as self-interaction cross-section are comparable with other operators considered, and hence we postpone analyzing this to future work.
\section{pseudoscalar mediator}
\label{pseudo}
In this section we will concern ourselves with pseudoscalar mediator denoted $a$. The nuclear coupling common to $\mathcal{O}^f_3$ and $\mathcal{O}^f_4$ is $g_{N} a \bar{N} \gamma_5 N$. These can be generated from quark couplings,
\begin{equation}
\mathcal{L} \supset g_{q_i} a \bar{q_i} \gamma_5 q_i.
\end{equation}

Limits on $g_{q_i} a \bar{q_i} \gamma_5 q_i$ from flavor physics are summarized in \cite{Dolan:2014ska}. For limit setting purposes, we consider the quark-Yukawa coupling scenario. These can be converted into limits on nucleons through \cite{Belanger:2008sj,Arina:2014yna}.

Finally constraints from supernovae cooling are derived for pseudoscalars in \cite{Chang:2018rso}.

\begin{figure}
\centering
\includegraphics[width=0.49\textwidth]{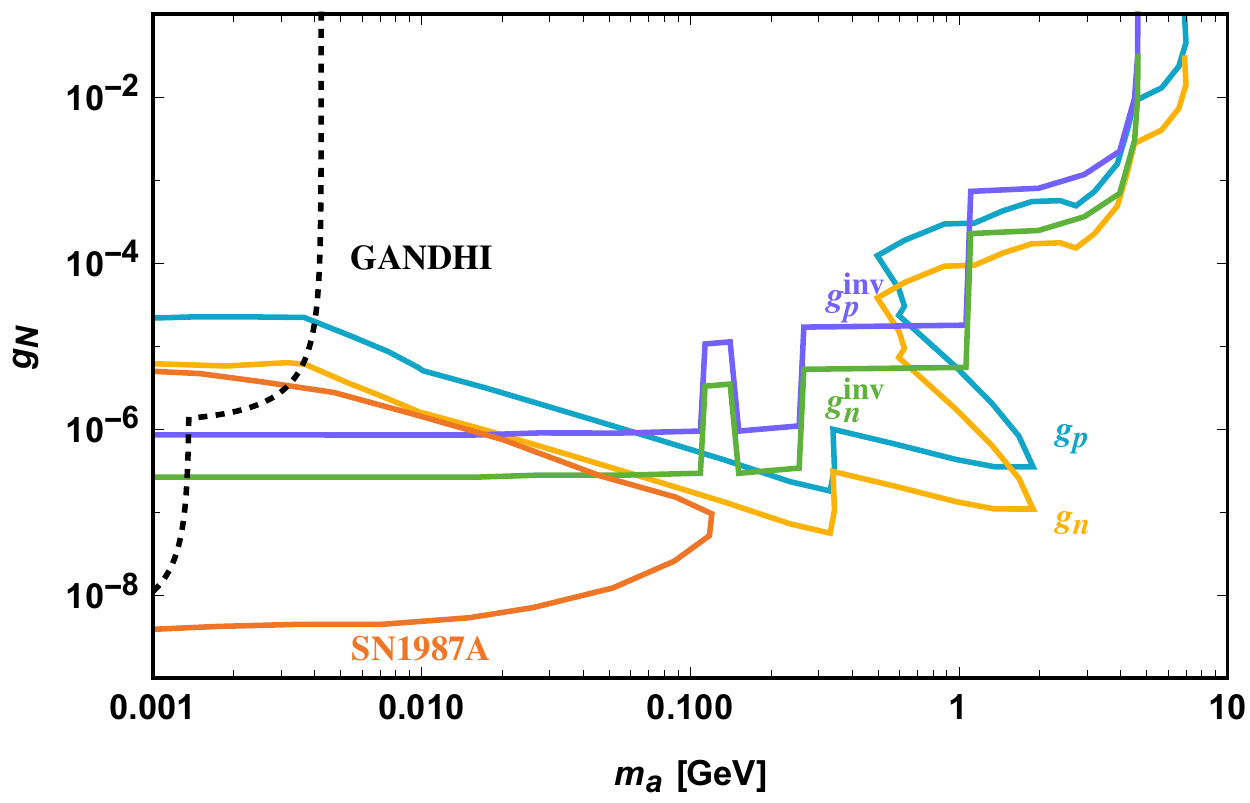}
\caption{Terrestrial and astrophysical limits on the pseudoscalar Yukawa coupling to nucleons are plotted as a function of pseudoscalar mass $m_a$. Terrestrial limits are adapted from \cite{Dolan:2014ska} and organized into proton and neutron couplings as well as into visible and invisible decay channels for the pseudoscalar. SN1987A constraints are from \cite{Chang:2018rso}. Also shown are projected limits from \cite{Benato:2018ijc}. }
\label{fig1}
\end{figure}

These constraints are summarized in Fig.~\ref{fig1}. We show constraints from meson decays both in the visible and invisible decay regimes for both neutron and proton couplings. Also shown are limits from supernova cooling. Interestingly, there is a pseudoscalar trapping window, between the supernova limits and the meson limits. Projected constraints from the proposed GANDHI experiment \cite{Benato:2018ijc} from nuclear decays through an invisible pseudoscalar could close this window upto $m_a=4\text{MeV}$. 

We will next summarize the DD and self-interaction cross-sections for pseudoscalar mediated operators.

\subsection{$\mathcal{O}^f_3$}

Starting with the Lagrangian,

\begin{equation}
\mathcal{L} \supset a (g_\chi \bar{\chi} \chi  + g_N \bar{N} \gamma_5 N)
\end{equation}

The transfer cross-section is given by \cite{Knapen:2017ekk}
\begin{equation}
\sigma_T^\text{born} \sim \frac{g_{\chi}^4}{2\pi m_\chi ^2 v^4} \{ \log(1+R^2)-R^2/(1+R^2)\}
\end{equation}
where $R=\frac{m_\chi v}{m_\phi}$, and the per-nucleon reference DD cross-section by,

\begin{equation}
\sigma_{\rm DD} = \frac{g_N^2 g_\chi^2}{32\pi}\frac{m_\chi^4 v^2}{m_N^2 (m_\phi^2+m_\chi^2 v^2)^2}
\end{equation}

\subsection{$\mathcal{O}^f_4$}
\begin{equation}
\mathcal{L} \supset a (g_\chi \bar{\chi}\gamma_5 \chi  + g_N \bar{N} \gamma_5 N)
\end{equation}
The transfer cross-section is given by,
\begin{equation}
\sigma_T^\text{born} =
\frac{g_\chi ^4 \left(6 \left(R^2+1\right) \log \left(R^2+1\right)+\left(R^4-3
   R^2-6\right) R^2\right)}{32 m_\chi^2 R^4 \left(R^2+1\right)}
 \end{equation}
\begin{equation}
\sigma_{\rm DD} = \frac{g_\chi^2 g_N^2}{192 \pi }\frac{m_\chi^4 v^4}{m_N^2 (m_\phi^2+m_\chi^2 v^2)^2}
\end{equation}

\section{Axial-vector Mediator}
As explained before, limits from $Z$ decays force us to consider a BSM axial-vector. Starting with the Lagrangian,
\begin{equation}
\mathcal{L} \supset g_N \bar{N} \gamma_\mu D^\mu \gamma_5 N
\end{equation}
Strict limits were derived in \cite{Dror:2017nsg} from the UV completion required to cancel anomalies on axial-vectors and will not be summarized here. There, explicit constraints were derived for currents coupled to right-handed quarks, but differ by an overall factor of 2 with the axial-vector case. Next, we consider operators generated by the axial-vector, $\mathcal{O}^f_8$ and $\mathcal{O}^v_6$.

\subsection{$\mathcal{O}^f_8$}
The DM interaction in this case is,
\begin{equation}
 \mathcal{L} \supset g_\chi \bar{\chi}\gamma_\mu D^\mu \gamma_5 \chi 
 \end{equation}
The transfer cross-section is given by, \cite{Knapen:2017ekk}
\begin{equation}
\sigma_T^\text{born} = \frac{6 g_{\chi}^4}{\pi m_\chi ^2 v^4} \{ \log(1+R^2)-R^2/(1+R^2)\}
\end{equation}
and finally, the DD cross-section is given by\cite{Freytsis:2010ne},

\begin{equation}
\sigma_{\rm DD} = \frac{12 g_N^2 g_\chi^2}{\pi} \frac{m_\chi^2}{(m_A^2 +m_\chi^2 v^2)^2}
\end{equation}
\subsection{$\mathcal{O}^v_6$}
The transfer cross-section is given by
\begin{align}
&\sigma_T^\text{born} =\nonumber \\&\frac{3 g_\chi ^4 \left(6 \left(R^2+1\right) \log \left(R^2+1\right)+\left(R^4-3
   R^2-6\right) R^2\right)}{32 m_\chi^2 R^4 \left(R^2+1\right)}
\end{align}
and finally, the DD cross-section is given by,

\begin{equation}
\sigma_{\rm DD} = \frac{3g_N^2 g_\chi^2}{2\pi} \frac{m_\chi^2 v^2}{(m_A^2 +m_\chi^2 v^2)^2}
\end{equation}

\section{Results}
\begin{figure*}[htpb]
\centering
\includegraphics[width=0.49\textwidth]{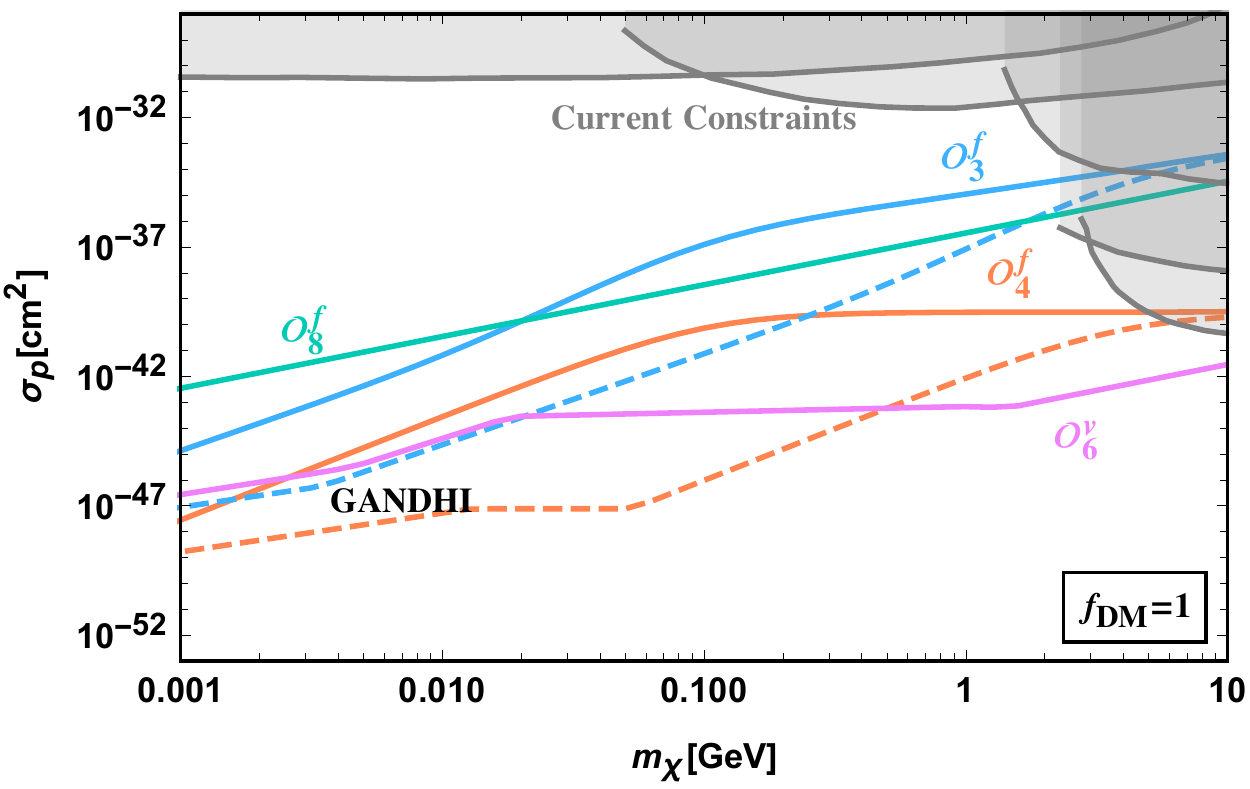}
\includegraphics[width=0.49\textwidth]{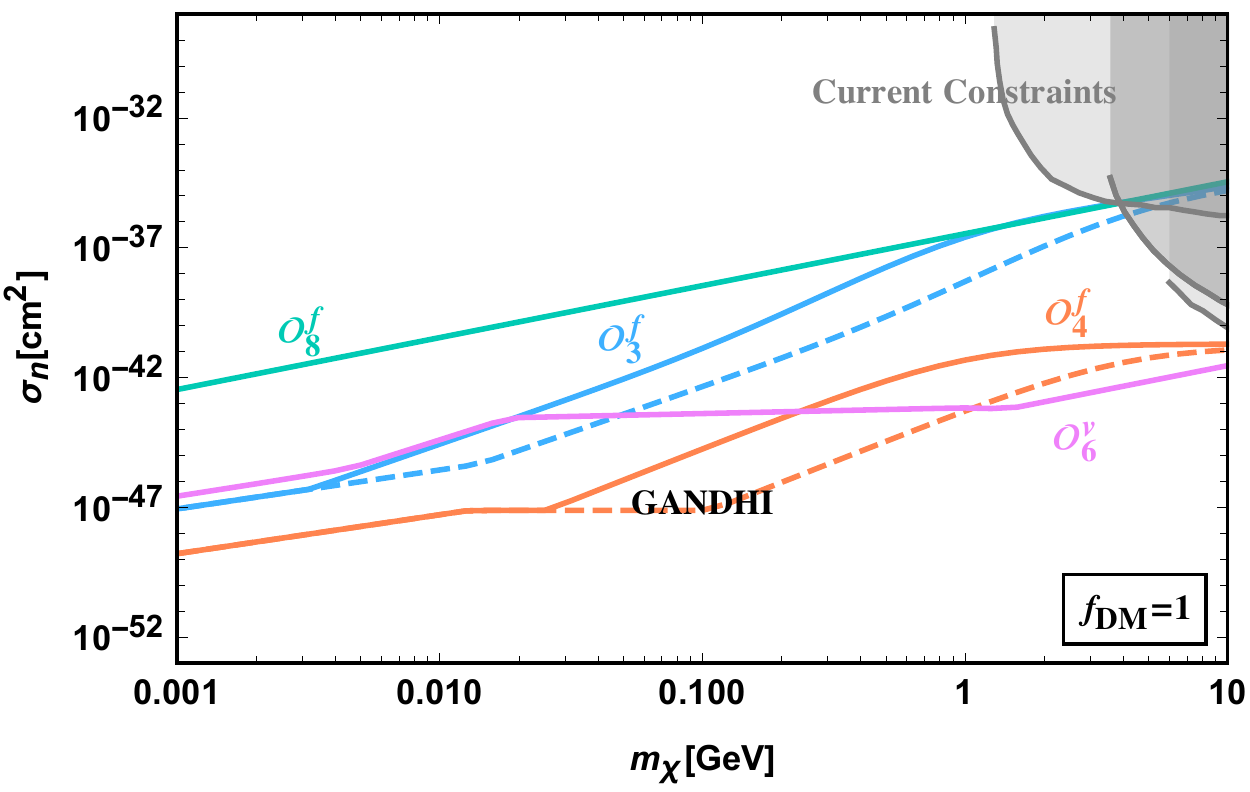}
\includegraphics[width=0.49\textwidth]{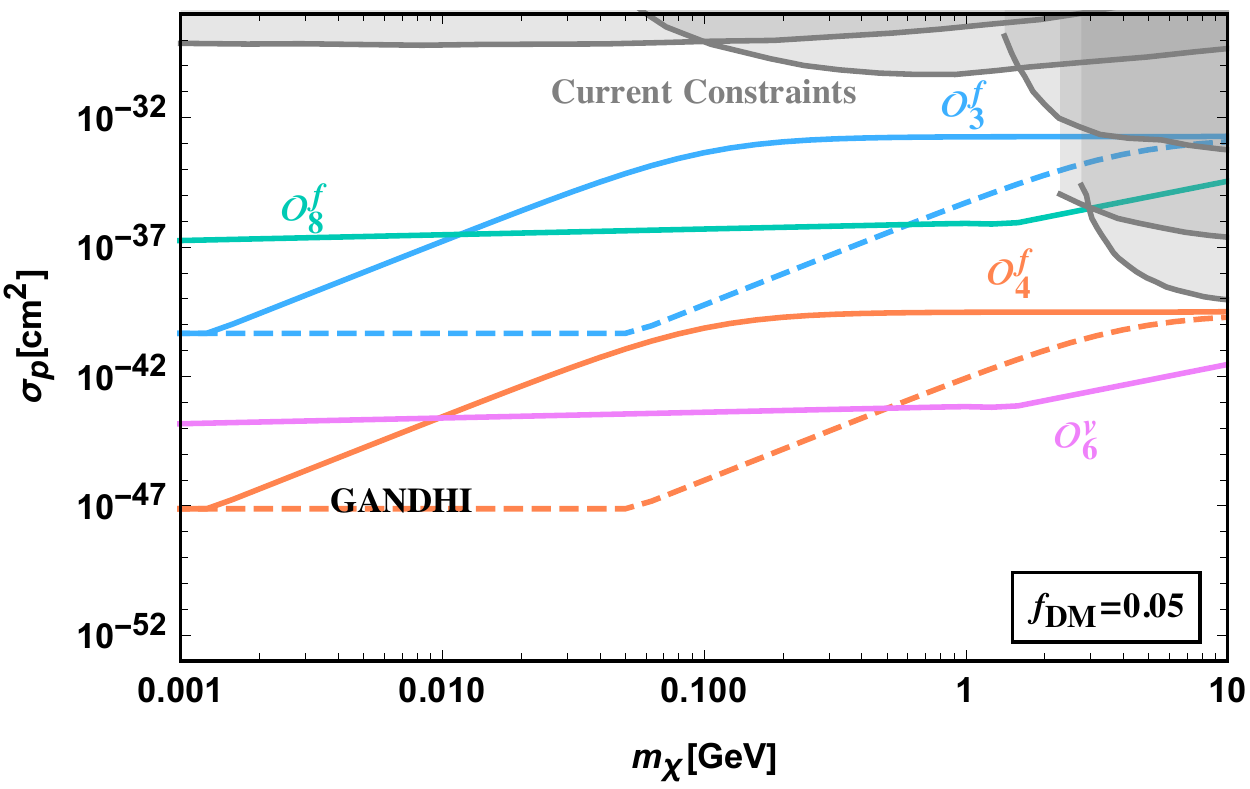}
\includegraphics[width=0.49\textwidth]{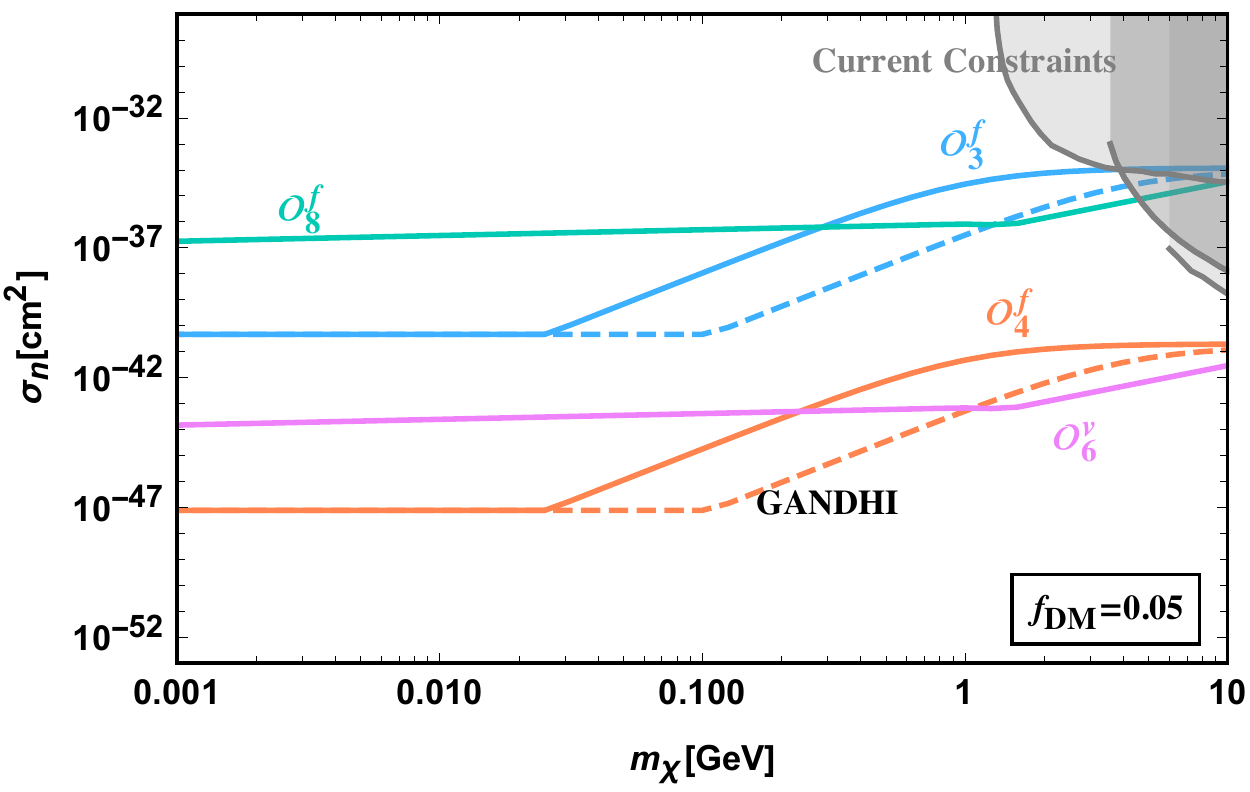}
\caption{Limits on per-nucleon reference DD cross-section for scattering with proton-spin ({\bf{Top-Left}}) and neutron-spin ({\bf{Top-Right}}) are shown as a function of DM mass $m_\chi$ for various operators considered in the text. Also shown are projected improvements (dotted lines) on constraints on the mediator from the GANDHI experiment \cite{Benato:2018ijc}. Current constraints from  \cite{Behnke:2016lsk,Amole:2017dex,Agnese:2017jvy,Collar:2018ydf,Bringmann:2018cvk} for $\sigma_p$ and \cite{Fu:2016ega,Akerib:2017kat,Agnese:2017jvy,Aprile:2019dbj} for $\sigma_n$ are also displayed. {\bf Bottom Panel}: Similar analysis but for sub-component DM where DM self-interactions are not constraining.}
\label{fig2}
\end{figure*}

Using the DD and self-interaction cross-sections in the previous sections, we set limits on the per-nucleon DD cross-section using the following procedure. If the DM species under consideration makes up all of DM, i.e. $f_{\rm DM}=1$, then the coupling of DM to the mediator, $g_{\chi}$ is constrained through the relevant self-interaction constraint from the Bullet Cluster system. We assume this cross-section to be 
\begin{equation}
\frac{\sigma_{T}^{\rm born}}{m_{\chi}} \le \frac{1 \text{cm}^2}{\rm gram} 
\end{equation}
Furthermore, the typical relative velocity in this system is assumed to be $v \sim 0.01$. If the DM species is subcomponent, $f_{\rm DM} \sim 0.05$, then the self-interaction constraints no longer hold. $g_{\chi}=1$ is instead imposed for perturbativity. 
Constraints on $g_N$, the SM-mediator coupling are obtained from the terrestrial and supernovae limits summarized in the previous sections. For the pseudoscalar coupling, we take the minimum of the visible and invisible channels for limit-setting purposes. 

With upper limits on both $g_{\chi}$ and $g_N$, one can obtain the maximum DD cross-section using the relevant formulae in the previous sections as a function of $m_\chi$, the DM mass and $m_a$/$m_A$, the mediator mass. Finally, we scan all relevant mediator mass to find the largest allowed $\sigma_{\rm DD}$ for a given $m_\chi$. This cross-section is plotted in Fig.~\ref{fig2} for scattering with protons, (top-left), with neutrons, (top-right) and for sub-component DM, (bottom-left) and (bottom-right). Different colors correspond to the 4 operators considered, $\mathcal{O}^f_3,\mathcal{O}^f_4,\mathcal{O}^f_8,\mathcal{O}^v_6$. Projected limits from the GANDHI experiment that can probe the pseudoscalar trapping window are also shown. Further, current DD constraints on SD proton scattering compiled from \cite{Behnke:2016lsk,Amole:2017dex,Agnese:2017jvy,Collar:2018ydf,Bringmann:2018cvk} and for neutron scattering from \cite{Fu:2016ega,Akerib:2017kat,Agnese:2017jvy,Aprile:2019dbj} are displayed.

For pseudoscalar mediated models, unsurprisingly $\mathcal{O}^f_3$ relaxes constraints compared to $\mathcal{O}^f_4$. Furthermore, constraints on $\sigma_n$ are stricter than $\sigma_p$ owing to the disparate factors involved in converting pseudoscalar quark couplings to pseudoscalar proton and neutron couplings. For axial-vector mediators, $\mathcal{O}^f_8$ sets better constraints compared to $\mathcal{O}^v_6$. Comparing to spin-independent limits derived in \cite{Knapen:2017ekk}, even the most pessimistic limits are tighter, owing to the tight constraints on the mediator in the case of axial-vectors and the momentum suppression in the DD cross-section in the case of pseudoscalar mediator. Also, unlike spin-independent cross-sections, limits do not relax around the GeV mass. 
Furthermore, relaxing the self-interaction constraints by looking at subcomponent DM, relaxes the DD cross-section limit differently for $\mathcal{O}^f_3$ and $\mathcal{O}^f_8$, with the latter having a much lower limit. This is because the pseudoscalar mediator also suppresses the self-interaction cross-section by powers of momentum exchange whereas for the axial-vector mediator no such suppression exists.

The neutrino floor for light dark matter experiments is target dependent and expected to be in the $\sigma_N\sim 10^{-47} - 10^{-41} \text{cm}^2$ range\cite{Green:2017ybv, Essig:2018tss}. Thus, this sets up a region of viable parameter space that can show signals solely in experiments sensitive to spin-dependent scattering, or to mediator searches like GANDHI.

\section*{Acknowledgements}
We are grateful to Jeff Dror, Zoltan Ligeti, Simon Knapen and Surjeet Rajendran for useful discussions. HR is supported in part by the DOE under contract DE-AC02-05CH11231. Some of this work was completed at the Aspen Center for Physics, which is supported by NSF grant PHY-1607611

\bibliography{biblio.bib}
\end{document}